\documentclass[aip,jcp,preprint]{revtex4-1}
\usepackage{amsmath}
\usepackage{amssymb}
\usepackage{graphicx}
\usepackage{subfig}
\newcommand{\mean}[1]{\langle#1\rangle}

\begin{document} 

\title{A many--body term improves the accuracy of effective potentials based on protein coevolutionary data}
\author{A. Contini}
\affiliation{Department of Physics, Universit\`a degli Studi di Milano, via Celoria 16, 20133 Milano, Italy}
\author{G. Tiana}
\email{guido.tiana@unimi.it}
\affiliation{Department of Physics, Universit\`a degli Studi di Milano, and INFN, via Celoria 16, 20133 Milano, Italy}
\date{\today}

\begin{abstract}
The study of correlated mutations in alignments of homologous proteins proved to be succesful not only in the prediction of their native conformation, but also in the developement of a two--body effective potential between pairs of amino acids. In the present work we extend the effective potential, introducing a many--body term based on the same theoretical framework, making use of a principle of maximum entropy. The extended potential performs better than the  two--body one in predicting the energetic effect of 308 mutations in 14 proteins (including membrane proteins). The average value of the parameters of the many--body term correlates with the degree of hydrophobicity of the corresponding residues, suggesting that this term  partly reflects the effect of the solvent. 
\end{abstract}

\maketitle

\section{Introduction}

The availability of simplified protein models with reduced degrees of freedom is useful for studying several biophysics problems. For example, the study of conformational changes in large protein systems is still unfeasable even on the fastest computers \cite{Piana:2013}. Conversely, with a reduced model it could be possible to study the thermodynamics of a 341--residues protein in a crowded environment \cite{Homouz:2008}. Free--energy differences upon mutation can be calculated {\it ab initio} only for small systems, while in more challenging cases one must resort to {\it ad--hoc} potentials \cite{Guerois:2002}. The elimination of solvent molecules is a standard example in which the use of a simplified model allows to study large and complex systems \cite{Arnarez:2015}. Anyway, the main problem associated with the reduction of the number of degrees of freedom in physical systems is the design of an effective potential, depending in a simple way on the remaining variables.

A way which has been followed several times to obtain effective potentials for proteins is the statistical approach \cite{Tanaka:1976,Miyazawa:1985,Mirny:1996}. The input data is the distribution of residues--residues contacts between the different types of amino acids in a selected set of proteins. One has to solve an inverse statistical--mechanics problem, searching for the potential which generated during natural evolution the frequencies of contacts which are actually observed in the selected set of proteins, assuming a Boltzmann relation between contact frequency and contact energy\cite{Shakhnovich:1993,Shakhnovich:1993uh,Tiana:2004ba}.

A variation of this approach is the calculation of contact energies based on the observed correlations between mutations in homologous proteins, using the same framework as that described in ref. \onlinecite{Morcos:2011jg} for a different problem, namely that is of predicting the native conformation of a protein from sequence information only. Here, pairs of residues which mutate in a correlated way in homologous sequence are regarded as in spatial contact, and from the full set of spatial contacts it could be possible to reconstruct the three--dimensional structure of several proteins. An inverse Ising--model formalism was used to subtract the effect of indirect correlations from the experimental data. 

The same formalism was then used in ref. \onlinecite{Lui:2013} to design an effective, non--portable two--body contact potential, assuming that the native conformation of the protein is known. This potential proved succesful in back--calculating residue--residue interactions in families of proteins generated by simulated evolution. It was also used to calculate the thermodynamic effect of mutations in four well-known proteins, giving correlation coefficients ranging between 0.65 and 0.89 between the experimental and the calculated $\Delta\Delta G$.

The formalism at the basis of refs. \onlinecite{Morcos:2011jg,Lui:2013} is meant to find the numerical values of the parameters of the effective energy
\begin{equation}
\mathcal{U}(\{\sigma_i\})=\sum_{i<j}e_{ij}(\sigma_i,\sigma_j)\Delta(|r_i-r_j|)+\sum_i h_i(\sigma_i),
\label{eq:u}
\end{equation}
from the knowledge of the observed frequencies $f_i(\sigma)$ of appearence of amino acid $\sigma$ at site $i$ and of the observed correlations $f_{ij}(\sigma,\tau)$ obtained in a set of $M$ aligned homologous sequences of length $L$. In Eq. (\ref{eq:u}), $\sigma_i$ is the type of residue at position $i$ of the protein, $\Delta(|r_i-r_j|)$ is a contact function which takes the value 1 if residues $i$ and $j$ are close in space (i.e., they contain a pair of heavy atom closer than a distance $d_r$) and zero otherwise,  $e_{ij}(\sigma_i,\sigma_j)$ is the interaction energy between residues $\sigma_i$ at position $i$ and $\sigma_j$ at position $j$, and $h_i(\sigma_i)$ is a one-body potential acting on each residue.

Once the numerical parameters entering Eq. (\ref{eq:u}) are calculated, the two--body energy $U(\{r_i\})=\sum_{i<j}e_{ij}(\sigma_i,\sigma_j)\Delta(|r_i-r_j|)$ can be applied for describing the conformational space of the protein. In ref. \onlinecite{Lui:2013}, for example, besides the calculation of mutational $\Delta\Delta G$, it was used to identify the frustrated regions of the protein. In doing so, the fields $h_i(\sigma)$ were regarded just as chemical potential meant to fix the average concentration of the twenty types of amino acids. Consequently, they were considered relevant only to control the underlying evolution of the set of homologous proteins, but not for the charcaterization of the conformational space of a well--defined sequence, of fixed amino--acid composition. Hence, they were neglected in the caluclation of the $\Delta\Delta G$.

However, one can think that the fields $h_i(\sigma)$ contain not only a chamical potential, but also a real interaction contribution associated with the position of a specific amino acid within the native conformation of the protein, not encoded in the two--body terms $e_{ij}(\sigma,\tau)$, and thus controlled by the $f_i(\sigma)$ rather than by the $f_{ij}(\sigma,\tau)$. This could be the case, for example, of the hydrophobic interaction, which depends, in first approximation, on the degree of burial of the $i$th site into the protein conformation, and not on the sum of two--body terms.

In the present work we want to disentangle the contribution to the potential which can be interepreted as an interaction term, from the one which is purely a chemical potential. We show that evolution of protein sequences onto a (fixed) native conformation can be described by an effective energy of the form  
\begin{align}
\mathcal{U}(\{\sigma_i\})&=\sum_{i<j}e_{ij}(\sigma_i,\sigma_j)\Delta(|r_i-r_j|)+\nonumber\\
&+\sum_i\eta_i(\sigma_i)+\sum_i\mu(\sigma_i),
\label{eq:u2}
\end{align}
where $\eta_i(\sigma_i)$ is the associated energy and $\mu(\sigma_i)$ is the chemical potential. We regard the first two terms as an effective interaction potential 
\begin{equation}
U(\{r_i\}) =\sum_{i<j}e_{ij}(\sigma_i,\sigma_j)\Delta(|r_i-r_j|)+\sum_i\eta_i(\sigma_i)\Theta_i(\{r_i\}),
\end{equation}
assigning a conformational dependence to its second term through a function $\Theta_i(\{r_i\})$ which measures the solvent--exposure of the $i$th residue.
We show that this effective potential predicts the experimental $\Delta\Delta G$ better than what the model involving only the two--body terms  did \cite{Lui:2013}.

\section{Derivation of the potential}

Given an alignment of $M$ homologous sequences, the input of the model is, as in the case of ref. \onlinecite{Morcos:2011jg}, the frequency $f_i(\sigma)$ of the amino acid of type $\sigma$ at site $i$ and the frequency $f_{ij}(\sigma,\tau)$ of the pair of types $\sigma$ and $\tau$ at sites $i$ and $j$, respectively, reweighted by the appropriate pseudocounts \cite{Altschul:2009} as
\begin{align}
&f_i(\sigma)=\frac{1}{M_{e}(x+y+z+1)}\times\nonumber\\
&\times\left[ \tilde{f_i}(\sigma)+x\frac{M_{e}}{q}+y\frac{\sum_j \tilde{f_j}(\sigma)}{L}+z \tilde{f_i}(\sigma) \right]\nonumber\\
&f_{ij}(\sigma,\tau)=\frac{1}{M_{e}(x+y+z+1)}\left[ \tilde{f_{ij}}(\sigma,\tau)+ x\frac{M_{e}}{q^2}+\right.\nonumber\\
&\left.+ \frac{y}{L^2M_{e}}\sum_{kl}\tilde{f_k}(\sigma)\tilde{f_l}(\tau)+\frac{z}{M_{e}}\tilde{f_i}(\sigma)\tilde{f_j}(\tau)  \right],   
\label{eq:emp}
\end{align}
where $\tilde{f_i}(\sigma)\equiv \sum_s \delta(\sigma,\sigma_i^s)/m_s$ and $\tilde{f_{ij}}(\sigma,\tau)\equiv \sum_s \delta(\sigma,\sigma_i^s)\delta(\tau,\sigma_j^s)/m_s$ are the raw frequencies,  $m_s$ is the number of sequences with similarity larger than 70\%, $q$ is the number of residue types and $M_{e}=\sum_s 1/m_s$ is an effective number of sequences. 

We shall search for a potential to generate a gobal distribution $p(\{\sigma_i\})$ for residue types in all the positions of the alignment, that matches the empirical distributions. In particular, we shall require that
\begin{align}
P(\tau)
\equiv&\sum_{\{\sigma_k\}}^{} \ p\big(\{\sigma_i\}\big) \sum_{i=1}^{L}\frac{\delta(\sigma_i,\tau)}{L}=  \frac{1}{L} \sum_i f_i(\tau)  \nonumber\\
\Delta P_i(\tau)
\equiv&\sum_{\{\sigma_k\}}^{} \ p\big(\{\sigma_i\}\big) \left[ \delta(\sigma_i,\tau) - \sum_{j=1}^{L}\frac{\delta(\sigma_j,\tau)}{L} \right] =\nonumber\\
=& f_{i}(\tau) - \frac{1}{L}\sum_j f_j(\tau) \nonumber\\
P_{ij}(\tau,\rho)
\equiv&\sum_{\{\sigma_k\}}^{} \ p\big(\{\sigma_i\}\big) \delta(\sigma_i,\tau) \delta(\sigma_j,\rho) =  f_{ij}(\tau,\rho).
\label{eq:freq}
\end{align}
The quantity $P(\tau)$ is the overall probability to find amino acid of type $\tau$ in any site, while $\Delta P_i(\tau)$ is the different between the probability in a specific site and the overall one, defined in such a way to be uncorrelated to $P(\tau)$. We also define the connected correlation function $C_{ij}(\tau,\rho)= f_{ij}(\tau,\rho)- f_i(\tau) f_j(\rho)$.

Since we have no other knowledge of the potential but the frequencies defined above, it seems reasonable to use the principle of maximum entropy with the constrains given by Eq. (\ref{eq:freq}) and the normalization condition of $p(\{\sigma_i\})$. Maximizing the entropy we obtain
\begin{equation}
\begin{split}
\label{eqn:new_prob1}
p\big(\{\sigma_i\}\big) = \frac{1}{\mathcal{Z}}\exp
		\Bigg[- &\sum_{i<j}^{} e_{ij}(\sigma_i,\sigma_j)
			- \sum_{i=1}^{L} \mu(\sigma_i) \\
			& - \sum_{i=1}^{L} \bigg( \widetilde{h}_i(\sigma_i) 
			- \frac{1}{L}\sum_{j=1}^{L} \widetilde{h}_i(\sigma_i) \bigg) 
		\Bigg] \ ,
\end{split}
\end{equation}
where the quantities $e_{ij}(\sigma,\tau)$, $\widetilde{h}_i(\sigma)$ and $\mu(\sigma)$ are  Langrange multipliers. Due to the formal similarity with Boltzmann's distribution, we  regard these quantities as effective energies. In particular, $\mu$ is site--independent and we assign to it the meaning of chemical potential. 

Assuming that there are $q$ types of amino acids, Eq. (\ref{eqn:new_prob1}) contains $q+Lq+q^2L(L-1)/2$ parameters. The experimental input of Eq. (\ref{eq:freq}) consists of $(q-1)+(L-1)(q-1)+(q-1)^2L(L-1)/2$ independent equations. Consequently, one has $1+(L+q-1)+(2q-1)L(L-1)/2$ free parameters which can be used to set the zeros of the energies. We must thus choose some $\bar{\sigma}$, $\widetilde{\sigma}$ and $\sigma^*$ such that
\begin{equation}
		\begin{split}
		\mu(\bar{\sigma})			&=0					\\
		\widetilde{h}_i(\widetilde{\sigma}) 	&=0		 \ \ \ \ \forall \, i 	\\
		\widetilde{h}_{\widetilde{i}}(\sigma)	&=0		 \ \ \ \ \forall \, \sigma 	\\
		e_{ij}(\sigma^*,\sigma) = e_{ij}(\sigma,\sigma^*)&=0	 \ \ \ \ \forall \, i,j,\sigma.
		\end{split}
		\label{eq:zeros}
\end{equation}
In other words, one has to choose an amino--acid type $\bar{\sigma}$ as the zero of the chemical potential, a type $\widetilde{\sigma}$ as the zero for the field $\widetilde{h}_i$ in each site (which in principle could be different from site to site), and a site $\widetilde{i}$ (the \emph{reference site}) in which the field $\widetilde{h}_{\widetilde{i}}(\sigma)=0$ for any type of amino acid.

For the purpose of determining the numerical values of the fields $ \widetilde{h}$ and of the chemical potentials $\mu$ in Eq. (\ref{eqn:new_prob1}), we follow the spirit of ref. \onlinecite{Morcos:2011jg} and write the argument of its exponential as an effective energy
\begin{equation}
		\label{eqn:new_hamA}
		\mathcal{U}_\alpha
		= \alpha \sum_{i<j}^{} e_{ij}(\sigma_i,\sigma_j)
		+ \sum_{i=1}^{L}\bigg[ 
			\mu(\sigma_i)
						+ \widetilde{h}_i(\sigma_i) 
			- \sum_{j=1}^{L} \frac{\widetilde{h}_i(\sigma_j)}{L} 
		\bigg]
\end{equation}
depending on the parameter $\alpha$ which controls the ratio between the two--body energy and the other energy terms. The associated Helmoltz free energy is
\begin{equation}
		\label{eqn:Fhelm}
		\mathcal{F}_\alpha = -\ln(\mathcal{Z}) = \mean{\mathcal{U}_\alpha} - S
\end{equation}
where temperature is immaterial in this derivation and is set to 1. The Gibbs free energy, obtained by a Legendre transform over the independent variables, is
\begin{align}
		\label{eqn:new_GibbsA_1}
		\mathcal{G}_\alpha &= \mathcal{F}_\alpha
		- L\sum_{\sigma=1}^{q-1}\mu(\sigma)  \frac{\partial [ - \ln (\mathcal{Z})]}{\partial  \mu(\sigma)}- \nonumber \\
		&- \sum_{i=1}^{L-1} \sum_{\sigma=1}^{q-1} \widetilde{h}_i(\sigma) \frac{\partial [ - \ln (\mathcal{Z})]}{\partial  \widetilde{h}_i(\sigma)},
\end{align}
in which the partial derivatives can be shown to be exactly $P(\sigma)$ and $\Delta P_i(\sigma)$, respectively. Consequently,
\begin{equation}
		\label{eqn:new_GibbsA_2}
		\mathcal{G}_\alpha = \mathcal{F}_\alpha
		- L\sum_{\sigma=1}^{q-1}\mu(\sigma) P(\sigma) - \sum_{i=1}^{L-1} \sum_{\sigma=1}^{q-1} \widetilde{h}_i(\sigma) \Delta P_i(\sigma).
\end{equation}
From Eq. (\ref{eqn:new_GibbsA_2}) it follows that the vaules of the fieds and of the chemical potentials can be obtained as
\begin{equation}
		\label{eqn:new_hb_fromg}
		\mu(\sigma)
		=-\frac{1}{L}\frac{\partial \mathcal{G}_\alpha}{\partial P(\sigma)} 
\end{equation}
\begin{equation}
		\label{eqn:new_ht_fromg}
		\widetilde{h}_i(\sigma)
		=-\frac{\partial \mathcal{G}_\alpha}{\partial \Delta P_i(\sigma)} 
\end{equation}

To find a manageable expression for $\mathcal{G}_\alpha$, this is expanded to the first order around $\alpha=0$, that is
\begin{equation}
		\label{eqn:GibbsAexpansion}
		\mathcal{G}_\alpha = \mathcal{G}_0
			 + \left. \frac{dG_\alpha}{d\alpha}\right|_{\alpha=0}  \cdot\alpha.
\end{equation}
In the zeroth--order term, the two--body energy does not appear because is proportional to $\alpha$, while the thermal average [cf. Eq. (\ref{eqn:Fhelm})] of the other three terms of the effective potential [cf. Eq. (\ref{eqn:new_hamA})] cancel out with the last two terms of Eq. (\ref{eqn:new_GibbsA_2}), leaving only the opposite of the entropy. Writing it in terms of the independent probabilities only, one obtains
\begin{equation}
		\begin{split}
		\label{eqn:New_G0}
		\mathcal{G}_0& 
		= \sum_{i=1}^{L-1} \sum_{\sigma=1}^{q-1} P_i (\sigma)\ \ln [  P_i (\sigma) ] + \\
		& + \sum_{i=1}^{L-1} \Bigg[ 	1 - \sum_{\sigma=1}^{q-1}  P_i (\sigma)	
		 \Bigg] \ln \bigg[1 - \sum_{\sigma=1}^{q-1} P_i (\sigma)\bigg] \\
		 & + \sum_{\sigma=1}^{q-1} \Bigg[ 		LP(\sigma) - \sum_{i=1}^{L-1}  P_i (\sigma)	
		 \Bigg] \ln \bigg[LP(\sigma) - \sum_{i=1}^{L-1}  P_i (\sigma)\bigg] \\
		  & + \Bigg[ 	1-\sum_{\sigma=1}^{q-1} \Big( LP(\sigma) - \sum_{i=1}^{L-1}  P_i (\sigma)\Big)	
		 \Bigg] \times\\
		 &\times\ln \bigg[1-\sum_{\sigma=1}^{q-1} \Big( LP(\sigma) - \sum_{i=1}^{L-1}  P_i (\sigma)\Big)\bigg].
		\end{split}
\end{equation}
In the second, third and fourth lines, the square brackets contains expressions for $P_i(\widetilde{\sigma})$, $P_{\widetilde{i}}(\sigma)$ and $P_{\widetilde{i}}(\widetilde{\sigma})$, respectively, which are not independent from the other probabilities  [cf. Eq. (\ref{eq:zeros})].

Remembering that $P(\sigma)+\Delta P_i(\sigma)=P_i(\sigma)$, the first--order term in Eq. (\ref{eqn:GibbsAexpansion}) results identical to that of ref. \onlinecite{Morcos:2011jg} and can be written as
\begin{equation}
		\begin{split}
		\label{eqn:New_G1}
		\left.\frac{d\mathcal{G}_\alpha }{d \alpha} \right|_{\alpha=0}
		= & \sum_{\sigma,\tau}\sum_{i<j}^{} e_{ij}(\sigma,\tau) P_i(\sigma) P_j(\tau)
		\end{split}
\end{equation}

Inserting into Eqs. (\ref{eqn:new_hb_fromg}) and (\ref{eqn:new_ht_fromg}) the expression of Eqs. (\ref{eqn:GibbsAexpansion}), (\ref{eqn:New_G0}) and (\ref{eqn:New_G1}), one obtains
\begin{equation}
 		\begin{split}	
		\label{eqn:New_htilde}
		 \widetilde{h}_m(\sigma) = -\ln\left[\frac{P_m(\sigma)}{P_m(\widetilde{\sigma})}\right] + \ln\left[\frac{P_{\widetilde{i}}(\sigma)}{P_{\widetilde{i}}(\widetilde{\sigma})}\right] - \\ 			-\alpha \sum_{\tau}\sum_{i | i\neq m}e_{mi}(\sigma,\tau) P_i(\tau)
		\end{split}
\end{equation}
and
\begin{equation}
 		\begin{split}	
		\label{eqn:New_hbar}
		 \mu(\sigma) = -\frac{1}{L}\sum_i^L\ln\left[\frac{P_m(\sigma)}{P_m(\bar{\sigma})}\right] 			
		 -\frac{\alpha}{L} \sum_{\tau}\sum_{i\neq m}e_{mi}(\sigma,\tau) P_m(\sigma)P_i(\tau).
		\end{split}
\end{equation}
On the other hand, since the the second term of Eq. (\ref{eqn:new_hamA}) can be written as $\sum_ih_i(\sigma_i)$, the two--body interaction terms do not change with respect to  ref. \onlinecite{Morcos:2011jg} [cf. Eq. (\ref{eq:u})], resulting in
\begin{equation}
e_{ij}(\sigma,\tau)=-C^{-1}_{ij}(\sigma,\tau).
\end{equation}

For sake of simplicity, we shall write the potential which controls the Boltzmann probability of Eq. (\ref{eqn:new_prob1}) as
\begin{equation}
		\label{eqn:new_hamB}
		\mathcal{U}
		=  \sum_{i<j}^{} e_{ij}(\sigma_i,\sigma_j)\Delta(|r_i-r_j|)
		+ \sum_{i=1}^{L}\eta_i(\sigma_i)+
			 \sum_{i=1}^{L}\mu(\sigma_i)	
\end{equation}
with  $\eta_i(\sigma)= \widetilde{h}_i(\sigma_i) - L^{-1}\sum_{j=1}^{L} \widetilde{h}_i(\sigma_j)$. The function $\Delta(|r_i-r_j|)$, which is zero if $|r_i-r_j|>d_r$, is also inserted in the potential to reduce the noise in the calculation of the energy in the native conformation. In fact, pairs of residues which do not interact directly whould have $e_{ij}=0$ due to the procedure described above to suppress indirect correlations. Effects such as the limited statistics of counts, or the approximation associated with the perturbative expansion of the potential could result in non-zero energies even in absence of direct correlations. Since we expect correlations to drop with the distance between residues, we introduce the $\Delta$ function (the choice of $d_r$ is discussed in detail in Sect. \ref{sect:parms}) to avoid spurious effects.

\section{Effect of the many--body term on the prediction of the experimental $\Delta\Delta G$}

To test the validity of the potential defined by Eq. (\ref{eqn:new_hamB}) we shall calculate the energetic effect $\Delta\Delta G$ of 308 point mutations on the stability of 14 proteins and compare them with the experimental values. 

The quantity  $\Delta\Delta G$ is the change in the difference between the free energies of the denatured and of the native state of the protein upon mutation. To calculate this quantity we need therefore to define the free energy of the denatured state. We assume, as often done when interpreting experimental data\cite{fersht}, that the mutation has no effect on the entropy of the chain, and that the interaction terms are zero in the denatured state (cf. Eq. \ref{eq:zeros}). Consequently, we shall make use of the interaction potential
\begin{equation}
U(\{r_i\})= \sum_{i<j}^{} e_{ij}(\sigma_i,\sigma_j)\Delta(|r_i-r_j|)
		+ \sum_{i=1}\Theta_i(\{r_i\})\eta_i(\sigma_i),
		\label{eq:u3}
\end{equation}
where $\Theta_i(\{r_i\})$ is some function of the coordinates of the protein which is 1 in the native conformation and zero in the denatured state. This function is not simply the sum of two--body terms (accounted by the first term of Eq. (\ref{eq:u3})), and consequently should be regarded as a many--body interaction. The chemical potential has been dropped because it plays no role in configurational space, in which the sequence $\{\sigma_i\}$ of the protein is fixed. The energetic effect of a point mutation is thus described by
\begin{align}
\Delta\Delta G(\sigma_i\to\sigma'_i)=& \sum_{j} [e_{ij}(\sigma_i,\sigma_j)  -  e_{ij}(\sigma'_i,\sigma_j)  ]\Delta(|r_i-r_j|)+\nonumber\\
		+& \eta_i(\sigma_i) -  \eta_i(\sigma'_i) .
\label{eq:ddg}
\end{align}

The protein--independent parameters of the model which gave the best results in terms of correlation coefficient between calculated and experimental  $\Delta\Delta G$ are $d_r=4.0$\AA, $\alpha=0.15$,   $x=0.5$, $y=0.1$, $z=1.0$ and the definition of the reference states as the most exposed site to the solvent occupied by polar or charged residues (for membrane proteins see below). The effect of variation of these parameters is described in Sect. \ref{sect:parms}.

\begin{figure}
\includegraphics[width=\linewidth]{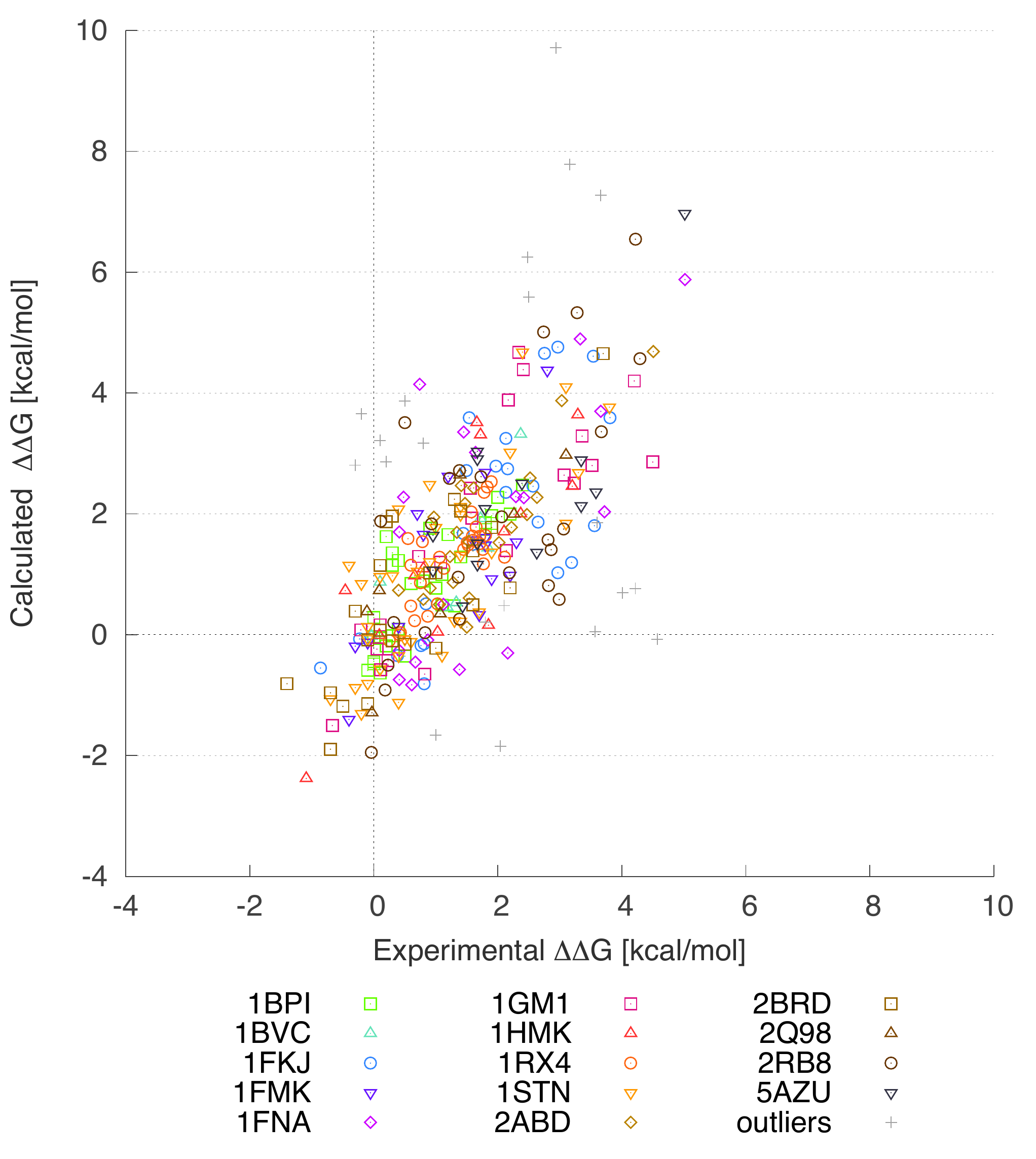}
\caption{The values of $\Delta\Delta G$ predicted by the model as a function of the corresponding experimental values.}
\label{fig:ddg}
\end{figure}

For this study we chose a set of protein domains with at least $1000$ homologs in the PFAM database, whose native structure is present in the PDB and on which the energetic effect of mutations has been characterized. This set is listed in Table \ref{tab:proteins}. The calculated values of $\Delta\Delta G$ is plotted versus their experimental values in Fig. \ref{fig:ddg}. The overall correlation coefficient between predicted and experimental values, excluding 23 outliers, is $r=0.77$. This should be compared with the value $r=0.47$ obtained predicting the $\Delta\Delta G$ making use of a potential including only the two body term $e_{ij}$, without the term $\eta_i$ (see Fig. S1 in the Supplemental Materials\cite{suppmat}).

A point is regarded as outlier if the difference between the calculated and experimental value is larger than $3\sigma$, where $\sigma$ is the error provided by the overall fit, also including the experimental error bars when available. Outliers can be classified into three cathegories (see Table S1 in the Supplemental Material\cite{suppmat}). 10 of them correspond to sites which are highly conserved, and consequently there is little (or no) statistics for the mutated sequence; 2 outliers are in sites which were experimentally characterized as structured in the denatured state, thus invalidating Eq. (\ref{eq:ddg}). The remaining 11 outliers cannot be explained in a satisfactory way, or the denatured state of their protein is not precisely experimentally determined.

The correlation coefficients between predicted and experimental data for each protein are displayed in Fig. \ref{fig:r} and are compared with those obtained without the term $\eta_i$ (cf. Fig. S2 in the Supplemental Materials\cite{suppmat} in which a detailed comparison of the $\Delta\Delta G$ is shown or each protein). 
We can see that including the new term  $\eta_i$ gives better correlation for most of the proteins (only 1BVC slightly decreases from 0.81 to 0.79 and 2ABD from 0.87 to 0.82).

In the set we have also a membrane protein (Bacteriorhodopsin, pdb entry 2BRD), for which this method is succesful in predicting $\Delta\Delta G$ for 24 mutations, without any outlier. To obtain this result we used a different reference state $\widetilde{i}$ than for cytosolic proteins, namely the most exposed hydrophobic site. Not unexpectedly, using for bacteriorhodobsin the same reference state used for the other solution proteins (i.e., the most exposed polar/charged site) gave a poor correlation coefficient of 0.53.

\section{Role of the parameters of the model} \label{sect:parms}

The model is defined by the values of $d_r$, $x$, $y$, $z$, and by the choice of the reference states in Eq. (\ref{eq:zeros}). Moreover, although the maximum--entropy principle is satisfied for $\alpha=1$, we found a better agreement with the experimental data for $\alpha<1$. Consequently, we regard $\alpha$ as a parameter of the model as well.

The dependence on the correlation coefficient $r$ between predicted and experimental $\Delta\Delta G$ on the interaction range $d_r$ of the two--body term is displayed in Fig. \ref{fig:dr} for some of the proteins studied above (see also Fig. S3 in  the Supplemental Material\cite{suppmat} for the other proteins). For all proteins $r$ is a decreasing function of $d_r$, modulated by an oscillating behavior. Its maximum lies between 3 and 6\AA, depending on the protein. The period of oscillation, of about 3--4\AA, is compatible with the size of the shells of other residues interacting with each residue in the native conformation. The best choice for $d_r$ seems to be 4.0\AA, although small variations of this have little effect in the prediction of the $\Delta\Delta G$.

The correlations coefficients $r$ as a function of $\alpha$ are displayed in Fig. \ref{fig:alpha} (cf. also S4 in  the Supplemental Material\cite{suppmat} ). Overall, they display a maximum at low values of $\alpha$ and decrease when $\alpha$ approaches 1. In few cases, the maximum is exactly at $\alpha=0$, that is when the terms $\widetilde{h}_i(\sigma)$ are decoupled from the terms $e_{ij}(\sigma,\tau)$ [see Eq. (\ref{eqn:New_htilde})]. In the production calculations we chose $\alpha=0.15$, although small variations of $\alpha$ have little effect if kept small, that is in the range where the perturbation expansion of the Gibbs free energy holds.

The coefficients $x$, $y$ and $z$ weight the pseudocounts, which are {\it a priori} probabilities meant to compensate the limited statistics in the alignments and make the correlation matrix invertible \cite{Altschul:2009,Morcos:2011jg,Lui:2013}. These three parameters weight the pseudocounts which depend, respectively, on the overall fraction of residue types, on the overall fraction of residue types in the specific position, and on the overall fraction of residue types in the specific pair of positions. The dependence of $r$ on these parameters is displayed in Fig. S5 in the Supplemental Material\cite{suppmat}. For most of the proteins the best choice is $x=0.5$, $y=0.1$, $z=1.0$.
Anyway, the quality of the results depends mainly on $z$, while the choice of $x$ and $y$ seems not critical.

While a natural and efficient choice for the reference state [see Eq. (\ref{eq:zeros})] of the two--body term $e_{ij}(\sigma,\tau)$ are the gaps in the alignment\cite{Lui:2013}, that for the reference state of the terms $\widetilde{h}_i(\sigma)$ is not straightforward. For cytosolic proteins, a sensible choice seems to be to set the reference site at the position of the most exposed polar or charged residue. The degree of solvent--exposure of a residue is quantified by the occupancy factor $S_{fact}$ defined in ref. \onlinecite{Gue:2002}. This choice assures that the many-body effective energy associated with the reference site does not change upon folding, since in the denatured state ($\Theta=0$) the sidechain is approximately as exposed as it is in the native state ($\Theta=1$). Suboptimal choices do not change dramatically the correlation coefficient, while the choice of hydrophobic sites significantly decreases it.

Bacteriorhodopsin, which is a membrane protein, behaves in the opposite way. Good results are obtained using as reference the most exposed hydrophobic site, which worsen choosing more hydrophilic sites.

\section{Properties of the $\eta$--term}

The term $\eta_i(\sigma)$ in the potential accounts for the contribution to the total energy which is not related to two--body interactions. As a result of the principle of maximum entropy, Eq. (\ref{eqn:new_prob1}) it is formally a one--body term of the potential, that is an external field. However, it is hard to justify an external filed in the present context, and consequently $\eta_i(\sigma)$ must be regarded as the result of the combined effect of the surrounding residues, that is a many--body term. 

\begin{figure}
\includegraphics[width=\linewidth]{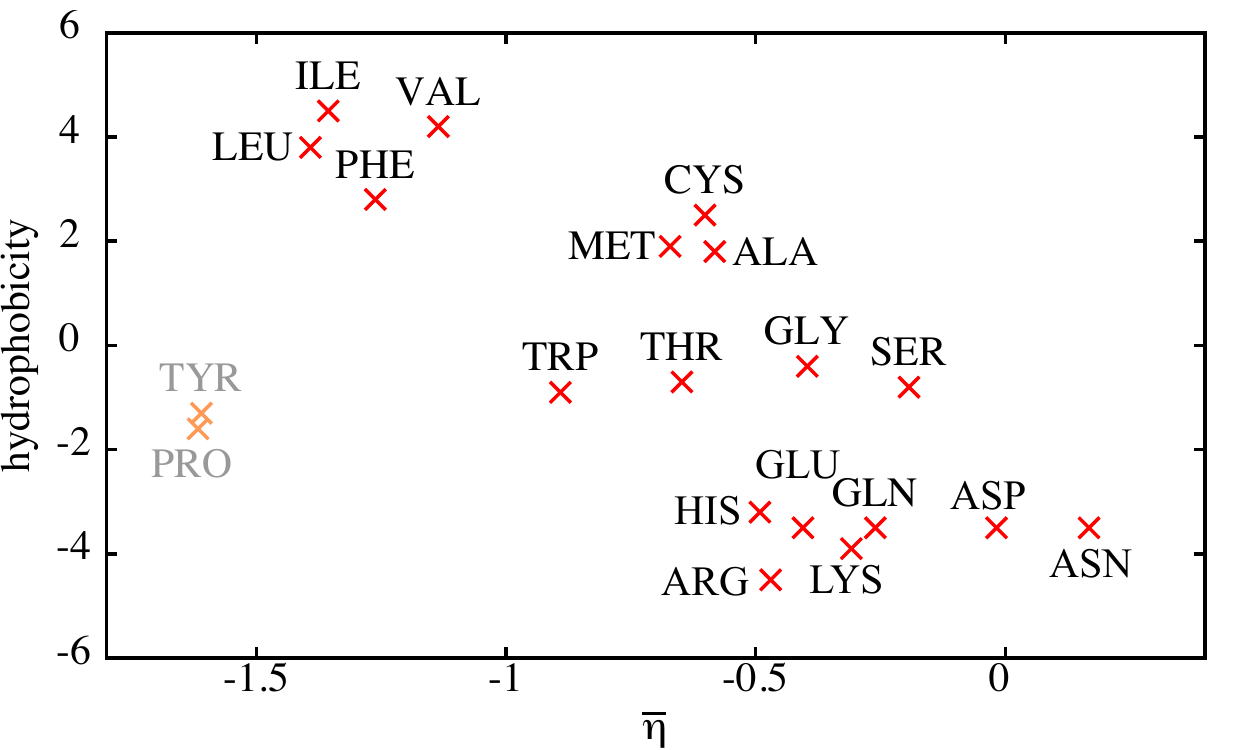}
\caption{The correlation between the average value of $\eta$ associated with each type of amino acid and its hydrophobicity, defined by the scale of Kyte and Doolittle. Excluding proline and tyrosine, the correlation coefficient is 0.81.}
\label{fig:hydro}
\end{figure}

The average value of $\eta$ over all its occurences in the proteins of Table \ref{tab:proteins} for each type of amino acid is displayed in Fig. \ref{fig:hydro}. Except that for proline and tyrosine, the average of $\eta$ has a good correlation ($r=0.81$) with the hydrophobicity of the corresponding residue, as measured by the scale of Kyte and Doolittle\cite{Kyte:1982}. This fact suggests that $\eta$ represents, at least partially, the contribution of the solvent to the positioning of the amino acids in the native conformation of the proteins. In fact, it is known that effective interaction associated with the presence of the solvent are intrinsically many--body\cite{delos:2000}.

While it is not completely unexpected that proline escapes the linear correlation between $\eta$ and hydrophobicity, because of its peculiar, rigid chemical structure, the behavior of tyrosine is surprising. Anyway, it cannot be explained in terms of poor statistics, since tyrosine appears in the proteins studied above with a frequency comparable to that of the other residues.

For the calculation of the $\Delta\Delta G$, the conformational dependence of the $\eta$--term of the potential has been regarded as two--state, in the sense that the only needed property of the function $\Theta_i(\{r_i\})$ in Eq. (\ref{eq:u3}) was to be 1 in the native state and 0 in the denatured state. To extend the use of the effective potential $U$ to characterize the conformational properties of a protein, one should define the full functional form of $\Theta_i(\{r_i\})$. The correlation of $\eta$--term with the hydrophobicity of the corresponding amino acids suggests that a reasonable assumption for $\Theta_i(\{r_i\})$ is the relative change in solvent exposure of the amino acid with respect to the native conformation, something which is indeed a many--body feature.

\section{Conclusions}

While effective potentials based on {\it ab initio} calculations contain no more and no less than the physical terms which are used in the underlying calculations, statistical potentials have the virtue to summarize all possible physical effects, even unknown ones. As an example of their power, statistical potentialls do not distinguish between globular and membrane proteins.  Moreover, their functional form is usually simpler, and then computationally cheaper, than other kinds of force fields. Thus, statistical potentials are potentially a powerful tool to study the properties of proteins. In particular,  those obtained from the analysis of mutational correlations proved efficient in predicting the native conformation of proteins\cite{Morcos:2011jg} and the experimental $\Delta\Delta G$\cite{Lui:2013}.

In the present work we have shown that the prediction of experimental $\Delta\Delta G$ can be further improved considering in the interaction potential a many--body term. This term arises naturally from a maximum--entropy principle, and can be parametrized within the same theoretical framework used for the two--body interaction term. It partially describes the effective interaction due to the solvent, but probably also other effects which cannot be reduced to a two--body interaction. As typical for statistical potentials, the choice of the reference state, that is the zero of the energy terms, plays a critical role in the correctness of the results.

\newpage

\begin{table}
\begin{tabular}{lccccc}
		\hline
		Protein/Domain	& Pdb	& Family 	& M & $M_{eff}$ & Mutat. \\
		\hline
		BPTI			& 1BPI		& 00014	& ${4915}$ & ${1566}$		& 35 \cite{yu1995contribution} 		\\
		Myoglobin		& 1BVC 	& 00042	& ${6000}$ & ${688}$		& 7  \cite{lin1993alpha}	 				\\
		FKBP1		& 1FKJ 	& 00014	& ${16739}$ & ${ 2284}$		& 26 \cite{main1998context} 		\\
		c-Src/SH3 dom.& 1FMK		& 00018	& ${10749}$ & ${1542}$		& 17 \cite{grantcharova1998important} \\
		Fibronectin/fnIII dom.	& 1FNA	& 00041	& ${17225}$ & ${8102}$		& 21 \cite{cota2000two}  			\\
		PTP-BL/PDZ dom.		& 1GM1	& 00595	& ${26099}$ & ${2715}$		& 23 \cite{gianni2007pdz} 			\\
		$\alpha$-Lactalbumin	& 1HMK	& 00062	& ${1035}$ & ${119}$			& 14 \cite{saeki2004localized}		\\
		ecDHFR			& 1RX4 	& 00186	& ${5237}$ & ${956}$			& 29 \cite{arai2005probing}			\\
		Staphiloc. nuclease	& 1STN 		& 00565	& ${4232}$ & ${1144}$		& 39 \cite{meeker1996contributions}	\\
		ACBP 		& 2ABD 	& 00887	& ${1677}$ & ${420}$			& 23 \cite{kragelund1999formation} 	\\
		Bacteriorhodopsin& 2BRD 	& 01036	& ${3174}$ & ${208}$			& 24 \cite{faham2004side}  			\\
		Del1-9-G129R-hPRL	& 2Q98 	& 00103	& ${1608}$ & ${97}$			& 9 \cite{keeler2009contribution}		\\
		Tenascin/fnIII dom.	& 2RB8 			& 00041	& ${17225}$ & ${8054}$			& 26 \cite{cota2000two}  			\\
		Azurin		& 5AZU  		& 00127	& ${1467}$ & ${282}$		& 15 \cite{wilson2005snapshots} 	\\
		\hline
		\end{tabular}
\caption{The list of protein domains, with the associated PDB structure the id of the PFAM family, the number $M$ of sequences in the family, the number $M_{eff}$ of effective sequences after reweighting for similarity and the number of mutations characterized experimentally.}
\label{tab:proteins}
\end{table}

\begin{figure}
\subfloat{\includegraphics[width=10cm]{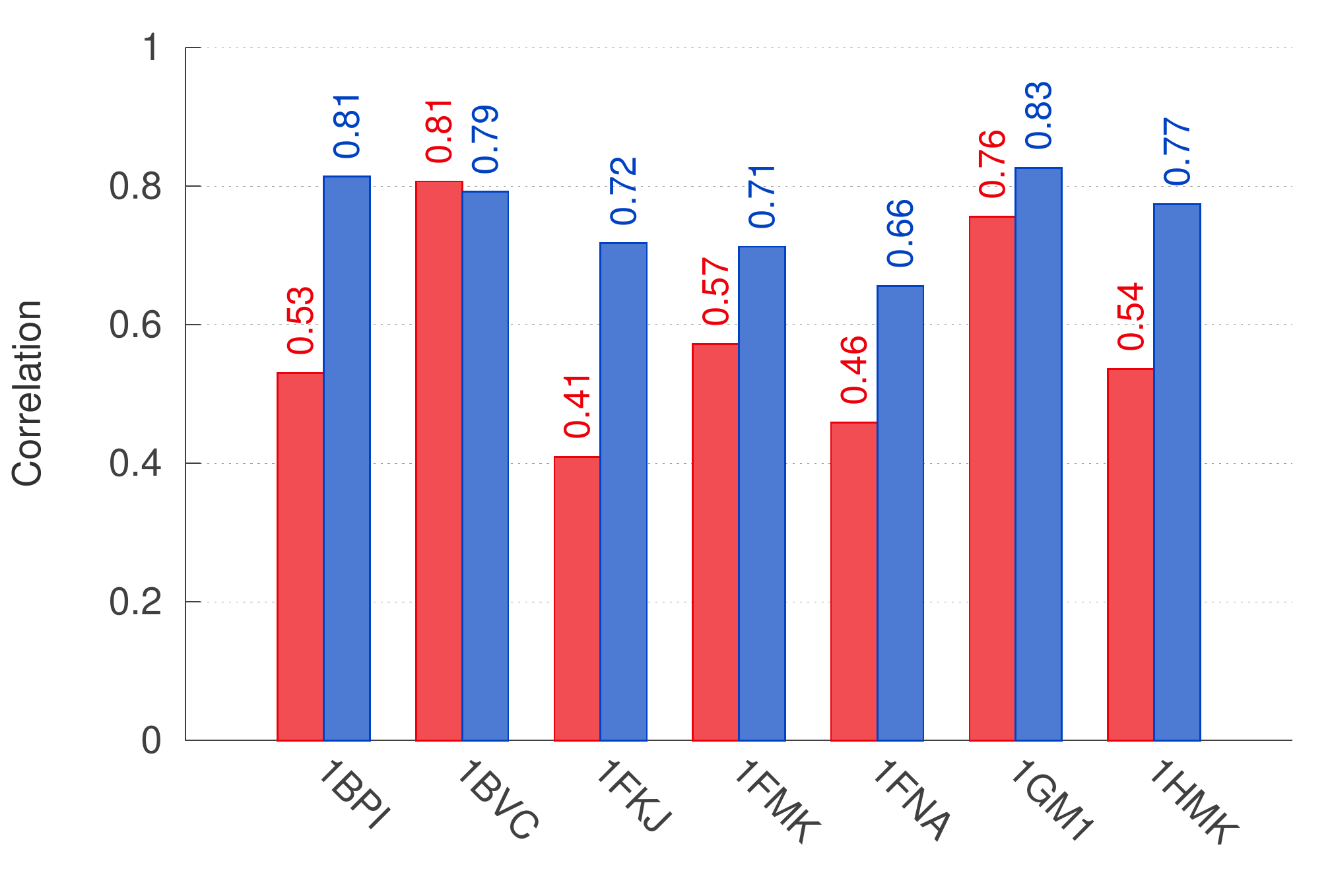}\label{fig:r1}}\\
\subfloat{\includegraphics[width=10cm]{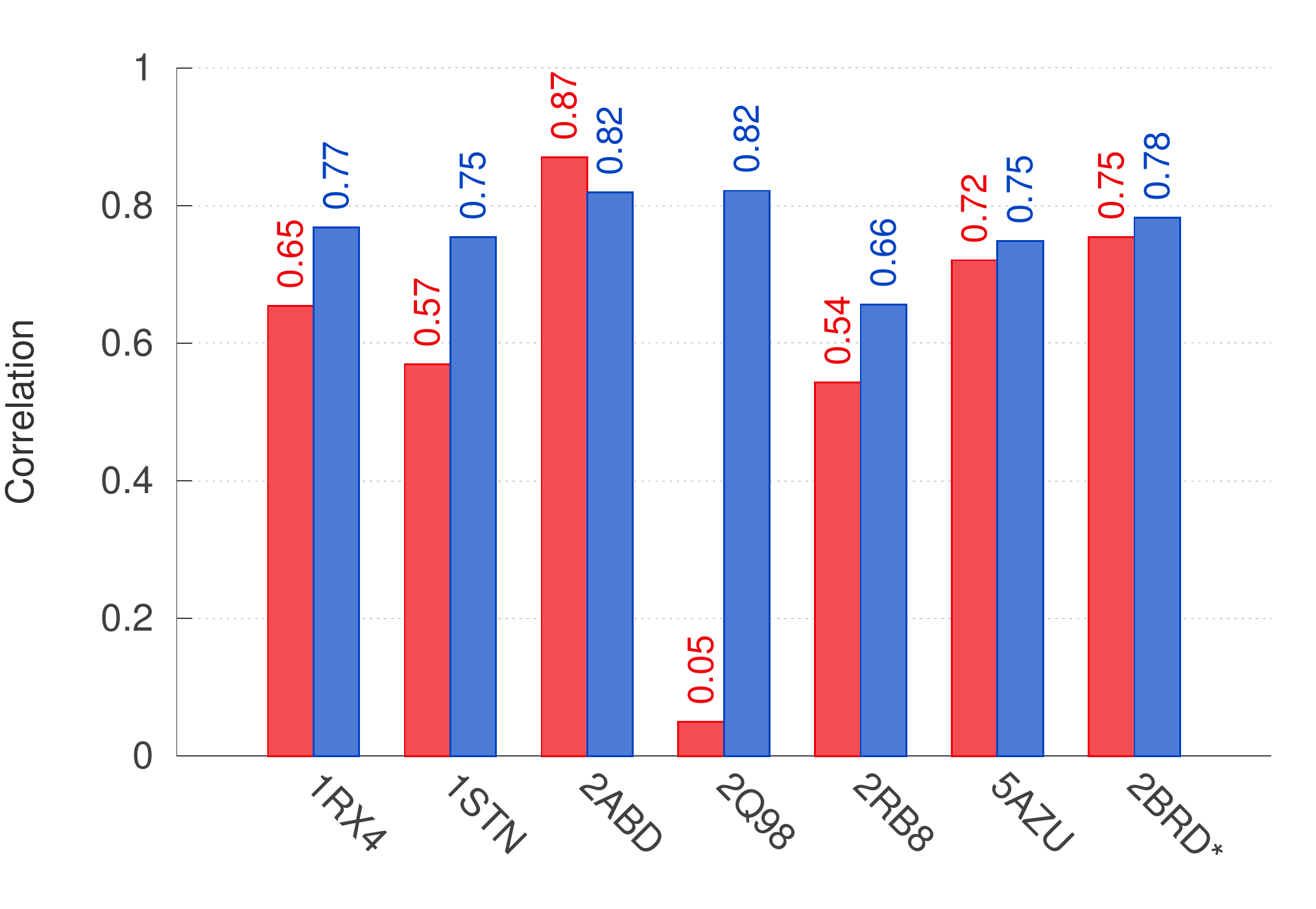}\label{fig:r2}}
\caption{The correlation coefficient between predicted and experimental $\Delta\Delta G$ for each protein. The red bars indicate the results obtained calculating the energies with the two--body term only, while the blue bars with the complete potential. The protein marked with an asterisk is a membrane protein.}
\label{fig:r}
\end{figure}

\begin{figure}
\includegraphics[width=\linewidth]{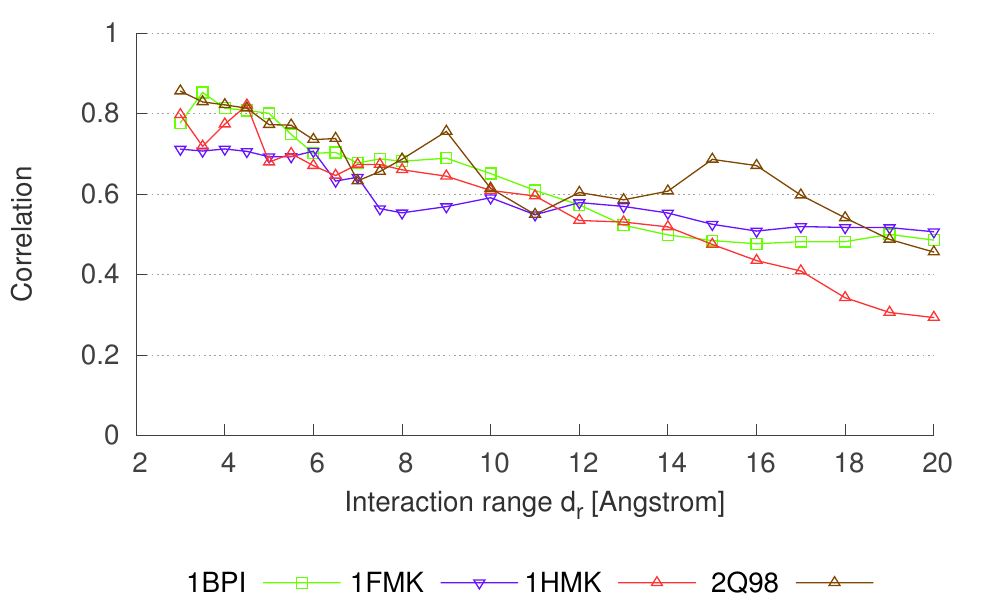}
\caption{The correlation coefficient $r$ as a function of the interaction range $d_r$ of the two--body energy term.}
\label{fig:dr}
\end{figure}

\begin{figure}
\includegraphics[width=\linewidth]{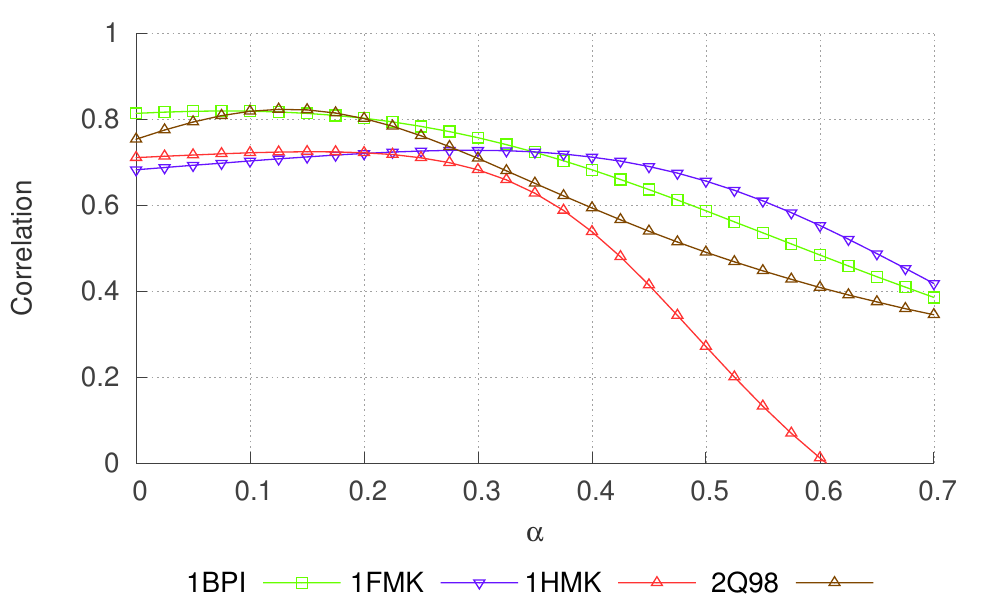}
\caption{The correlation coefficient $r$ as a function of the perturbation coefficient $\alpha$.}
\label{fig:alpha}
\end{figure}

\begin{figure}
\includegraphics[width=\linewidth]{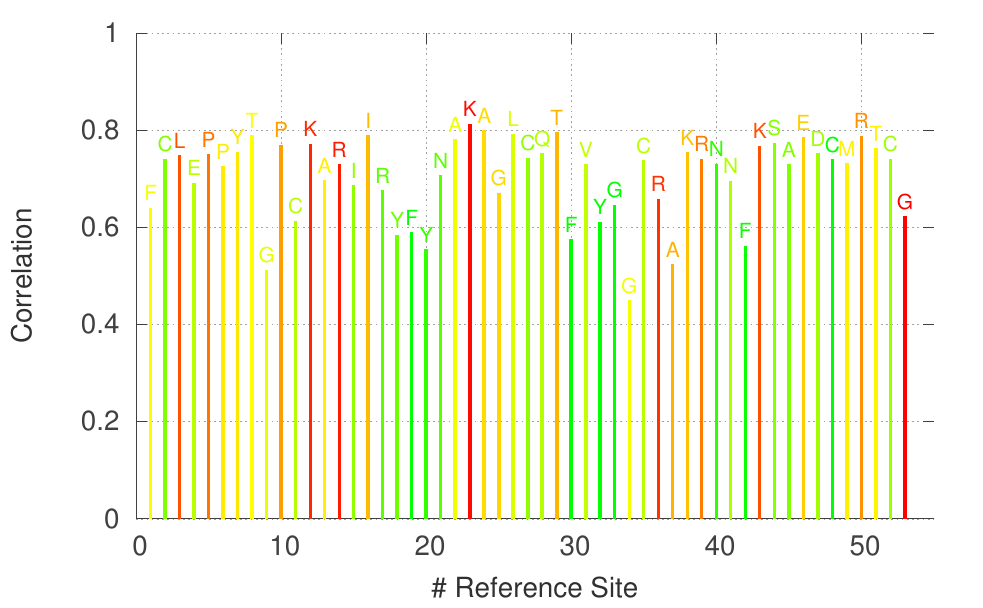}
\caption{The correlation coefficient $r$ as a function of the choice of the reference state for $\widetilde{h}_i(\sigma)$ for 1BPI. The color code indicates the degree of solvent exposure. The color scale goes from red (exposed) to green (buried). Residue K23 (K26 according to the numbering of the pdb) is selected as the reference state.}
\label{fig:reference}
\end{figure}

\end{document}